\documentclass[12pt]{JHEP3}

\usepackage{epsfig}
\usepackage{graphicx}
\usepackage{subfigure}
\usepackage{bm}

\newcommand{\beq}{\begin{equation}}
\newcommand{\eeq}{\end{equation}}
\newcommand{\baq}{\begin{eqnarray}}
\newcommand{\eaq}{\end{eqnarray}}
\newcommand{\fnl}{f_{\mathrm{NL}}}
\newcommand{\gnl}{g_{\mathrm{NL}}}
\newcommand{\rdec}{r_{\mathrm{dec}}}

\title{Non-Gaussian Fingerprints of Self-Interacting Curvaton}

\author{Kari Enqvist$^{a,b,}\footnote{E-mail: kari.enqvist@helsinki.fi}$,
Sami Nurmi$^{c,}\footnote{E-mail: s.nurmi@thphys.uni-heidelberg.de}$,
Olli Taanila$^{a,b,}$\footnote{E-mail: olli.taanila@iki.fi},
Tomo Takahashi$^{d,}\footnote{E-mail: tomot@cc.saga-u.ac.jp}$ \\
$^a$Physics Department, FIN-00014 University of Helsinki, Finland;\\
$^b$Helsinki Institute of Physics, FIN-00014 University of Helsinki, Finland;\\
$^c$Institute for Theoretical Physics, University of Heidelberg, 69120 Heidelberg, Germany;\\
$^d$Department of Physics, Saga University, Saga 840-8502, Japan }

\abstract {We investigate non-Gaussianities in self-interacting curvaton models treating both renormalizable and non-renormalizable polynomial interactions. We scan the parameter space systematically and compute numerically the non-linearity parameters $\fnl$ and $\gnl$. We find that even in the interaction dominated regime there are large regions consistent with current observable  bounds. Whenever the interactions dominate, we discover significant deviations from the relations $\fnl\sim \rdec^{-1}$ and $\gnl\sim \rdec^{-1}$ valid for quadratic curvaton potentials, where $\rdec$ measures the curvaton contribution to the total energy density at the time of its decay. Even if $\rdec\ll 1$, there always exists regions with $\fnl\sim 0$ since the sign of $\fnl$ oscillates as a function of the parameters. While $\gnl$ can also change sign, typically $\gnl$ is non-zero in the low-$\fnl$ regions. Hence, for some parameters the non-Gaussian statistics is dominated by $\gnl$ rather than by $\fnl$. Due to self-interactions, both the relative signs of $\fnl$ and $\gnl$ and the functional relation between them is typically modified from the quadratic case, offering a possible experimental test of the curvaton interactions.}


\preprint{HIP-2009-33/TH\\HD-THEP-09-32}

\begin{document}

\section{Introduction}

In the curvaton scenario \cite{curvaton}, primordial perturbations
originate from quantum fluctuations of a curvaton field $\sigma$ which
remains effectively massless during inflation and contributes very
little to the total energy density. After the end of inflation, the
curvaton energy density stays nearly constant until the field becomes
effectively massive, while the dominating radiation component
generated at reheating scales as $\rho_{\rm rad} \propto a^{-4}$. The
curvaton contribution to the total energy density therefore grows
rapidly after the end of inflation and its perturbations start to
affect the expansion history. In this way the initial isocurvature
fluctuations of the curvaton field get gradually converted into
adiabatic curvature perturbations. The observed primordial
perturbation can originate solely from the curvaton perturbations,
although scenarios with mixed inflaton and curvaton perturbations are
also possible \cite{mixed}. As the curvaton finally decays and the
decay products thermalize, the adiabatic hot big bang epoch is
recovered and the curvature perturbation freezes to a constant value
on superhorizon scales.

Predictions of the curvaton scenario depend crucially on the form of
the curvaton potential \cite{Dimopoulos:2003ss,kesn,kett,Enqvist:2008be,Huang:2008zj,Kawasaki:2008mc,Chingangbam:2009xi,Enqvist:2009zf,Chambers:2009ki} (and the background evolution \cite{Enqvist:2009eq}).
  Although the
curvaton must be weakly coupled to the thermal bath after the end of
inflation, interactions of some type should be present in realistic
models.

In the present paper, we consider self-interacting curvaton models
with the generic potential %%
\beq
 \label{V}
 V=\frac{1}{2}m^2\sigma^2+\lambda\sigma^{n+4}\ ,
 \eeq
 with $n=0,2,4,6$
 and $\lambda>0$. We set $M_{\rm P}=(8\pi G)^{-1/2}\equiv1$ throughout
 the paper. In \cite{Enqvist:2009zf} it was found numerically that the
 predictions of such models can deviate significantly from the
 extensively studied quadratic case \cite{LUW}, as shown already in \cite{kesn,kett} in the limit of small interactions.
 For non-quadratic potentials the amplification
 of curvaton energy after inflation is less efficient than in the
 quadratic case, but one may still generate the observed amplitude of
 primordial perturbation since the curvaton perturbation generated during inflation,
 $\delta\sigma_{*}/\sigma_{*}$, can be larger than $10^{-5}$. As shown
 in \cite{Enqvist:2009zf}, the correct amplitude can be achieved in a
 relatively large part of the parameter space.

 However, the dynamics
 is very complicated and, for non-renormalizable potentials, the
 curvature perturbation $\zeta=\Delta N$ (here understood to contain
 all orders of perturbation theory) displays oscillations as a
 function of the initial curvaton field value $\sigma_{*}$. This
 reflects the dynamics of transition from curvaton oscillations in the
 non-quadratic part of the potential to the quadratic part, as
 discussed in \cite{Enqvist:2009zf}. For the (marginally)
 renormalizable case, $n=0$, the transition does not give rise to an
 oscillatory behaviour of $\zeta$ but nevertheless affects the final
 value of the perturbation.

 Here we extend the analysis of \cite{Enqvist:2009zf} and
 focus on primordial non-Gaussianities generated in the class
 of curvaton models with the potential (\ref{V}). We use the $\Delta
 N$ formalism and solve the equations of motion numerically. We
 perform a systematic scan through the parameter space and evaluate
 the non-linearity parameters $\fnl$ and $\gnl$. Combining this
 information with the amplitude of perturbations, we find the
 regions of the parameter space that are consistent with current
 observational constraints. As a result of the oscillatory behaviour
 of $\zeta(\sigma_{*})$, the observational constraints on the
 self-interacting model (\ref{V}) differ significantly from the
 quadratic case.

 The paper is organized as follows. In Sect.~\ref{nongausscurv} we give our basic definitions and discuss the
 generic features of the non-linearity parameters $f_{\rm NL}$ and
 $g_{\rm NL}$. In Sect.~\ref{quartic}, we tackle the special
 case of the renormalizable, four point interaction, for which we can
 use the analytical estimates derived in \cite{Enqvist:2009zf}. In
 Sect.~\ref{nonrenormpot}, we extend our analysis to
 non-renormalizable interactions with $n=2,4,6$ by resorting to
 numerics. This section contains our main results. Finally, in Sect.~\ref{yakyak} we present our conclusions.

\section{Non-Gaussianities in self-interacting curvaton models}

\label{nongausscurv}

We use the $\delta N$ formalism \cite{deltaN,recent_deltaN} and assume
that the curvature perturbation arises solely from perturbations of a
single curvaton field generated during inflation. The curvature
perturbation $\zeta$ can then be expanded as
\beq
\label{zeta}
\zeta(t,{\bm x}) = N'(t,t_{*})\delta\sigma_{*}({\bm x})
+\frac{1}{2}N''(t,t_{*})\delta\sigma_{*}({\bm x})^2
+\frac{1}{6}N'''(t,t_{*})\delta\sigma_{*}({\bm x})^3 \cdots \, .
\eeq
Here $N(t,t_{*})$ is the number of e-foldings from an initial
spatially flat hypersurface with fixed scale factor $a(t_*)$ to a
final hypersurface with fixed energy density $\rho(t)$, evaluated
using the FRW background equations. The final time $t$ should be chosen as some
(arbitrary) time event after the curvaton decay. The prime denotes a
derivative with respect to the initial curvaton value
$\sigma_{*}$. Here we take $t_{*}$ to be some time during inflation
soon after all the cosmologically relevant modes have exited the
horizon and assume the curvaton perturbations $\delta\sigma_{*}$ are
Gaussian at this point. The expansion (\ref{zeta}) is then of the form
\beq
 \label{fnl_gnl}
 \zeta(t,{\bm x})
 =
 \zeta_{\rm g}(t,{\bm x})
 +\frac{3}{5}f_{\rm NL}\zeta_{\rm g}(t,{\bm x})^2
 +\frac{9}{25}g_{\rm NL}\zeta_{\rm g}(t,{\bm x})^3+\cdots \ .
\eeq
where $\zeta_{\rm g}(t,{\bm x})$ is a Gaussian field and the
non-linearity parameters are given by
\begin{eqnarray}
\label{fnl_def}
\fnl &=& \frac{5}{6}\frac{N''}{N'^2}\\
\label{gnl_def}
\gnl &=& \frac{25}{54}\frac{N'''}{N'^3}\ .
\end{eqnarray}
Here we neglect all the scale dependence of the non-linearity parameters \cite{Byrnes:2009pe}. With this assumption and neglecting higher order perturbative corrections, the constants $\fnl$ and $\gnl$ measure the amplitudes of the three- and four-point correlators of $\zeta$ respectively.

We assume the curvaton obeys slow roll dynamics during inflation and
introduce a parameter $r_{*}$ to measure its contribution to the total
energy density $\rho$ at $t_{*}$
\beq
 \label{r_star}
 r_{*}=\frac{\rho_{\sigma}}{\rho}\Big|_{t_{*}}\simeq\frac{V(\sigma_{*})}{3H_{*}^2}\ll 1\ .
\eeq
The inflationary scale $H_{*}$ appears as a free parameter in our
analysis, up to certain model dependent consistency
conditions. Assuming inflation is driven by a slowly rolling inflaton
field, we need to require $H_{*}\ll 10^{-5} \sqrt{\epsilon}$ in order
to make the inflaton contribution to $\zeta$ negligible. In this setup
we also need to adjust $\epsilon=-\dot{H}_{*}/H_{*}^2$, determined by
the inflaton dynamics, to give the correct spectral index,
$n-1=2\epsilon-2\eta_{\sigma\sigma}$ \cite{Wands:2002bn}. The curvaton
contribution, $\eta_{\sigma\sigma}=V''(\sigma_{*})/3H_{*}^2$, is
typically negligible because of the subdominance of the curvaton. The curvaton mass needs to be small but the same also holds for the inflaton mass.
%The curvaton mass needs to be small but the same also holds for the inflaton mass. The difference between these two fields is that onee is an isocurvature direction, i.e., subdominant.
% is supposed to be very small compared to $H_\ast$.
We assume this minimal setup in the current work for definiteness since
our main goal is to discuss the new features arising from curvaton
self-interactions.

After the end of inflation, we assume the inflaton decays completely
into radiation and the universe becomes radiation dominated. We
introduce a phenomenological decay constant $\Gamma$ to account for
the coupling between the radiation and the curvaton component. The
evolution of the coupled system is given by
\begin{eqnarray}
\label{frw1}
&&
\ddot{\sigma} + (3H+\Gamma)\dot{\sigma}+m^2\sigma + \lambda(n+4)\sigma^{n+3} = 0\\
\label{frw2}
&&
\dot{\rho_\mathrm{r}} = -4H\rho_\mathrm{r} + \Gamma \dot{\sigma}^2\\
\label{frw3}
&&
3H^2 = \rho_\mathrm{r} + \rho_\sigma \ .
\end{eqnarray}
The initial conditions are given by $\rho_{\mathrm{r}} = 3H_*^2$ and
$\rho_\sigma = V(\sigma_*) = r_* / (1-r_*) \rho_{\mathrm{r}}$ specified at time $t_*$ corresponding to the end of inflation. We also put
$\dot{\sigma} = 0$. Given the parameters $n$, $\lambda$ and $m$, which
determine the potential (\ref{V}), and the two initial conditions
$r_*$ and $H_*$, we can calculate $N$ in (\ref{zeta}) from this set of
equations. To find the curvature perturbation, we set
$\delta\sigma_{*}=H_{*}/(2\pi)$ and compute
$\zeta=N(\sigma_{*}+\delta\sigma_{*})-N(\sigma_{*})$. For a given set
of parameters, we adjust the decay width $\Gamma$ so that the observed
amplitude is obtained, $\zeta \sim 10^{-5}$ \cite{wmap}.

We treat $\Gamma$ as a free parameter since we have not specified
the curvaton couplings to other matter, in particular to the Standard
Model fields. However, since the primordial perturbations have been
observed to be adiabatic to a high degree \cite{wmap}, the curvaton
should decay before dark matter decouples in order not to produce
isocurvature modes. Assuming the freeze-out temperature for
an LSP type dark matter model with
$T_{\rm DM}\sim~{\cal O}(10)$ GeV, this translates to a rough bound
\begin{equation}
\label{eq:isocurvature}
\Gamma \gtrsim 10^{-15} \mathrm{GeV} = 10^{-33} \ .
\end{equation}
While this bound could be relaxed in non-minimal constructions, we
assume it for definiteness for the rest of our work.

\subsection{Analytical considerations}

In Sect.~\ref{nonrenormpot} we solve the set of equations
(\ref{frw1}) - (\ref{frw3}) numerically and compute the resulting
curvature perturbation (\ref{zeta}) and the non-linearity parameters
(\ref{fnl_def}) and (\ref{gnl_def}). However to gain some physical
insight of the results, it is useful to start by discussing generic
approximative analytic expressions that can be derived for $\fnl$ and
$\gnl$. Assuming the curvaton decays instantaneously \cite{sudden} at $H_{\rm
  dec}=\Gamma$ and neglecting the coupling between curvaton and
radiation before $t_{\rm dec}$, one can can estimate the non-linearity
parameters by \cite{kett,LR,Sasaki:2006kq}
\baq
\label{f_nl}
\fnl&=&\frac{5}{3\rdec}\left(1+\frac{\sigma_{\rm osc}\sigma_{\rm osc}''}{(\sigma_{\rm osc}')^2}\right)-\frac{5}{3}-\frac{5\rdec}{8}\\
\label{g_nl}
\gnl&=&
\frac{25}{54}\left[ \rule{0pt}{4ex}\right.\frac{4}{\rdec^2}\left(\frac{\sigma_{\rm osc}^2\sigma_{\rm osc}'''}{(\sigma_{\rm osc}')^3}+\frac{3 \sigma_{\rm osc} \sigma_{\rm osc}''}{(\sigma_{\rm osc}')^2}\right)-\frac{12}{\rdec}\left(1+\frac{\sigma_{\rm osc}\sigma_{\rm osc}''}{(\sigma_{\rm osc}')^2}\right)+ \\
\nonumber &&\frac{1}{2}\left(1+\frac{\sigma_{\rm osc}\sigma_{\rm osc}''}{(\sigma_{\rm osc}')^2}\right)+\frac{30\rdec}{4}+\frac{27 \rdec^2}{16}\left]\rule{0pt}{4ex}\right.\ .
\eaq
Here $\rdec=\rho_{\sigma}/\rho|_{\rm dec}\sim \sigma_{\rm
  osc}^2(m/\Gamma)^{1/2}$ and $\sigma_{\rm osc}$ is the envelope of
the oscillating curvaton field at some time $t_{\rm osc}$ after the
beginning of oscillations in the quadratic part of the potential. The
results are independent of the precise choice of $t_{\rm osc}$.

If the curvaton potential is exactly quadratic, $\sigma_{\rm osc}
\propto  \sigma_{*}$ and the non-linearity parameters $\fnl$
and $\gnl$ are uniquely determined by $\rdec$, i.e. by the curvaton
energy density at the time of decay. However, interactions in general
make the function $\sigma_{\rm osc}(\sigma_{*})$ non-linear and, especially in
the limit $\rdec \ll 1$, the quadratic predictions can be greatly
altered due to the derivative terms in (\ref{f_nl}) and
(\ref{g_nl}). If the interactions dominate the curvaton dynamics at
the time of inflation, the non-linearity parameters can only very
crudely be approximated by $f_{\rm NL}\sim \rdec^{-1}$ and $g_{\rm NL}\sim \rdec^{-2}$. As will be discussed in Sect.~\ref{nonrenormpot}, in the interacting case the non-linearity parameters can deviate significantly from these na\"{i}ve estimates. This follows from the non-trivial behaviour of
derivatives of $\sigma_{\rm osc}(\sigma_{*})$ in (\ref{f_nl}) and
(\ref{g_nl}).

In particular, we find that for $n=2,4$ in the potential (\ref{V}) the signs of the
non-linearity parameters $f_{\rm NL}$ and $g_{\rm NL}$ oscillate as
a function of $\sigma_{*}$. Therefore, even if $\rdec \ll 1$, it is
always possible to find regions where the non-Gaussianities are
accidentally suppressed and do not conflict current observational
bounds. For example, from (\ref{f_nl}) we see that $f_{\rm NL}\sim
0$ for $\rdec \ll 1$ whenever ${\sigma_{\rm osc}\sigma_{\rm
    osc}''}/{(\sigma_{\rm osc}')^2}\sim -1$. Using the definitions
(\ref{fnl_def}) and (\ref{gnl_def}) we find,
\beq
\label{eq:fnlprime}
 f'_{\rm NL}(\sigma_{*})=\frac{9}{5}N'\left(g_{\rm NL}-\frac{4}{3}f_{\rm NL}^2\right)\, ,
\eeq
which gives an estimate $|\Delta \sigma|\sim (\gnl N')^{-1}$ for the
typical size of the regions $f_{\rm NL}\sim 0$. This is larger than
the range of field values probed by the curvaton fluctuations produced
during inflation, $|\delta\sigma_{*}|\sim 10^{-5}/N'$, if $g_{\rm
  NL}\lesssim 10^{5}$. Therefore, for $\gnl\lesssim 10^{5}$ we may
conclude that the regions $\fnl\sim 0$ correspond to a self-consistent
choice of initial conditions and are not destabilized by the
fluctuations produced during inflation.

\section{Renormalizable potential}

\label{quartic}

The special case of a quartic interaction term in the potential,
\beq
\label{quarticpot}
V=\frac 12 m^{2}\sigma^{2}+\lambda\sigma^4~,
\eeq
can be discussed using the analytical estimates presented in
\cite{Enqvist:2009zf}. If the interaction dominates over the quadratic
term at the time of inflation, the curvaton oscillations start in an
effectively quartic potential. As the amplitude of the oscillating
field decreases, the quartic term dilutes away and the potential
becomes quadratic. Assuming the universe remains radiation dominated
until the curvaton decay, $\rdec\ll 1$, the envelope of the
oscillating curvaton in the quadratic regime is given by
\beq
\label{sigma_as}
\sigma_{\rm as}(t)\simeq\frac{\sigma_{\rm osc}(\sigma_{*})}{(mt)^{\frac{3}{4}}}\, .
\eeq
The function $\sigma_{\rm osc}(\sigma_{*})$ can be estimated using
Eq. (4.9) in \cite{Enqvist:2009zf}, which assumes
$\lambda\sigma_{*}^2\gtrsim m^2$ and is derived to leading order
precision in $\rdec\ll1$. In approaching the quadratic limit,
$\lambda\sigma_{*}^2\ll m^2$, the analytical estimates of
\cite{Enqvist:2009zf} can no longer be used to obtain quantitative
results but the qualitative features remain correct.

\begin{figure}[!h]
\centering
\includegraphics[width=11cm, height=5 cm]{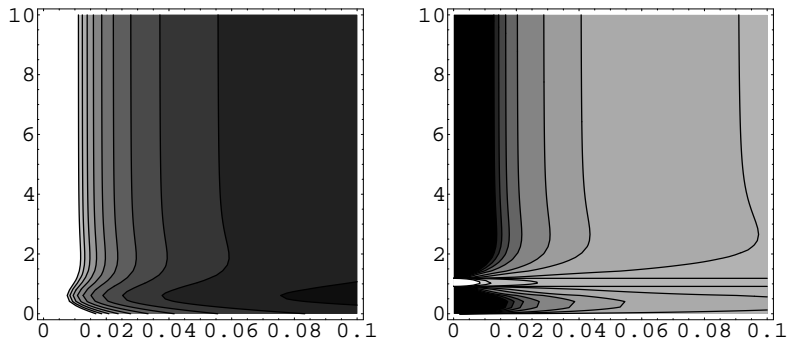}
\caption{Contour plots of $f_{\rm NL}$ (left panel) and $g_{\rm NL}$
  (right panel) with variables $\rdec$ and
  $\sqrt{\lambda}\sigma_{*}/m$ as $x$ and $y$ axes, respectively. On
  the left panel, the contours run from $0$ (black) to $100$ (white)
  with spacing of $10$. On the right panel, they run with spacing of
  $500$ from $-5000$ (black) to $0$ (white).}
\label{fig:fnl_and_gnl}
\end{figure}

To leading order in $\rdec$, Eqs. (\ref{f_nl}) and (\ref{g_nl})
read
\baq
\label{f_nl_small_r}
\fnl&\simeq&\frac{5}{3\rdec}\left(1+\frac{\sigma_{\rm osc}\sigma_{\rm osc}''}{(\sigma_{\rm osc}')^2}\right)\\
\label{g_nl_small_r}
g_{\rm NL}&\simeq&\frac{50}{27\rdec^2}\left(\frac{\sigma_{\rm osc}^2\sigma_{\rm osc}'''}{(\sigma_{\rm osc}')^3}+\frac{3 \sigma_{\rm osc} \sigma_{\rm osc}''}{(\sigma_{\rm osc}')^2}\right)\ ,
\eaq
and by substituting (\ref{sigma_as}) into these, we find
(semi)analytical predictions for $f_{\rm NL}$ and $g_{\rm NL}$. The
results are illustrated in Fig. (\ref{fig:fnl_and_gnl}).

Although the $n=0$ case does not result into oscillatory behaviour of
$\zeta(\sigma_{*})$, non-monotonous features appear when considering
derivatives of $\zeta(\sigma_{*})$, i.e. the coefficients $N'', N'''$
etc. in (\ref{zeta}). These features become the more pronounced the
higher the order of the derivative $N^{(m)}$ is. This can also be
observed in Fig. (\ref{fig:fnl_and_gnl}) where both $\fnl$ and $\gnl$
display non-monotonous behaviour as a function of $\sigma_{*}$ in the
region $\lambda\sigma_{*}^2\sim m^2$. For $f_{\rm NL}$ the
non-monotonous features are very mild but $g_{\rm NL}$ shows a
significantly more sensitive dependence on $\sigma_{*}$.

As discussed in \cite{Enqvist:2009zf}, for $\lambda\sigma_{*}^2\sim
m^2$ the transition from the quartic to the quadratic part of the
potential takes place soon after the onset of oscillations when
$\sigma$ and $\dot{\sigma}$ still act as independent degrees of
freedom. These quantities in general do not have a similar dependence
on the initial field value $\sigma_{*}$. Variations of $\sigma_{*}$
therefore affect the effective equation of state of the oscillating
curvaton in a non-trivial fashion and this is the origin of the
structure seen in Fig. (\ref{fig:fnl_and_gnl}). If
$\lambda\sigma_{*}^2\gg m^2$ or $\lambda\sigma_{*}^2\ll m^2$, no
similar structure is seen. In the former case, the transition into
quadratic region happens when the dynamics of the oscillating curvaton
is already well described by a single dynamical degree of freedom, the
amplitude $\sigma$.  In the latter case the potential is almost
Gaussian from the beginning.

\section{Non-renormalizable potentials}
\label{nonrenormpot}

The discussion in the preceding section, together with the oscillatory
behaviour of $\zeta$ presented in \cite{Enqvist:2009zf}, leads to an
expectation that for non-renormalizable curvaton potentials the nai\"{i}ve estimates $\fnl\sim 1/r_{\rm dec}$ and $\gnl \sim
\fnl^2$ can be violated for a range in the parameter space. It is
however not obvious at all how large this range might be. Since the analtyical approximations used in the previous section cannot be applied to non-renormalizable potentials, we use numerical methods to track the dynamics at hand.

To calculate the values of $\fnl$ and $\gnl$ for the non-quartic
cases, we use code developed for \cite{Enqvist:2009zf} to compute the
values numerically using Eqs.~\ref{fnl_def} and \ref{gnl_def}. To
obtain the derivatives, we calculate $N$ for five different initial
conditions separated by a spacing $a$, and then use five-point stencil
to calculate the first, the second and the third derivative of $N$. We
adjust the stepping $a$ so that numerical noise is
minimized. Furthermore, we adjust $a$ for each pixel independently.

\subsection{Qualitative features and differences from na\"{i}ve expectations}

For all choices of $m$ and $n$, the parameter space is divided into
two areas by a line where the quadratic and the non-quadratic terms
are equal for the initial curvaton field value, i.e.,
\[ \frac{1}{2}m^2\sigma_*^2 = \lambda \frac{\sigma_*^{n+4}}{M^n}\, .\] Below this
quadratic line we should recover the behaviour predicted analytically
for the quadratic case. Above this line the self-interactions modify
the behaviour substantially.

For $n=6$ case the behaviour in the non-quadratic regime is smooth and
qualitatively in close resemblance with the case $n=0$, which was
described in the previous section. This is due to the fact that if
$n\geq6$, the field does not oscillate in the non-renormalizable part
of the potential at all. However, for the cases $n=2$ and $n=4$ large
oscillations of $\zeta$ as a function of its parameters and initial
conditions are present in the non-quadratic regime as described in
\cite{Enqvist:2009zf}. Here the expressions $\fnl \sim 1/\rdec$ and
$\fnl^2 \sim \gnl$ provide only very rough estimates of the
non-linearity parameters, and in the interaction dominated region the
actual values of $\fnl$ and $\gnl$ can deviate significantly from
these estimates. This behaviour can be traced back to the terms
including derivatives of $\sigma_{\mathrm{osc}}$ in the Eqs.
(\ref{f_nl}) and (\ref{g_nl}).

\begin{figure}[!ht]
\centering \subfigure[$|\fnl|$ plotted against the intial conditions $r_*$ and $H_*$.] {
\includegraphics[width=7 cm]{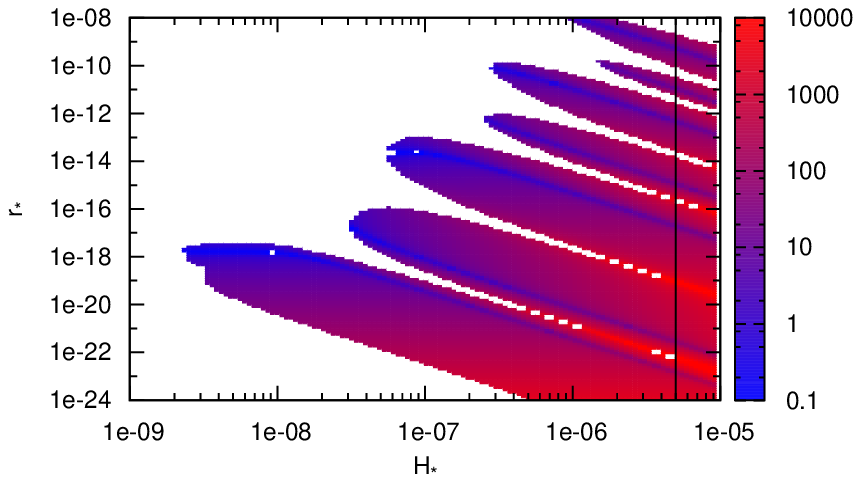}
\label{fig:fnl_r_contour}} \subfigure[$|\fnl|$ plotted against $\rdec$ for a fixed value $H_* = {5\times10^{-6}}$, corresponding to the vertical stripe in Fig.~\protect\ref{fig:fnl_r_contour}. Red points correspond to $\fnl > 0$ and blue points to $\fnl < 0$.] {
\includegraphics[width=7 cm]{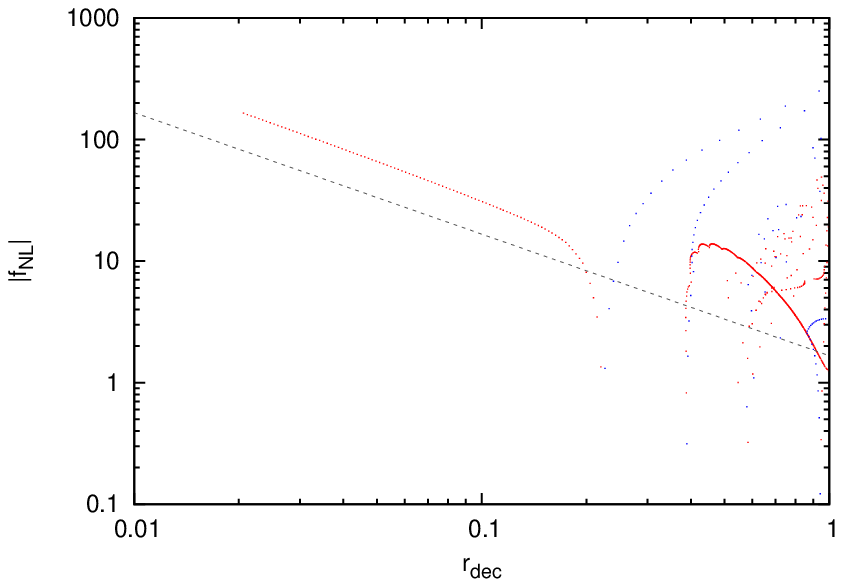}
\label{fig:fnl_r_stripe}}
\subfigure[$|\fnl|$ plotted against $\rdec$ for all values of $H_*$ in \protect\ref{fig:fnl_r_contour}. Red points correspond to $\fnl > 0$ and blue points to $\fnl < 0$.]
{
\includegraphics[width=7 cm]{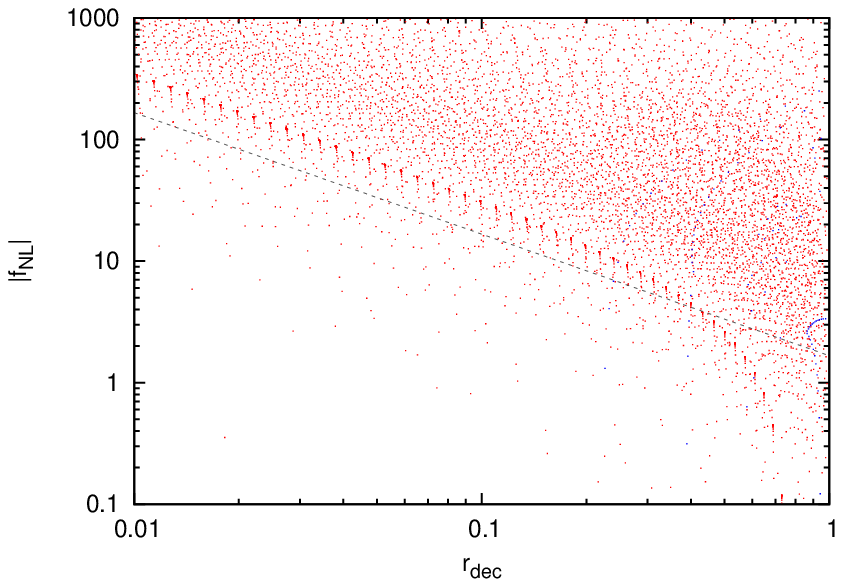}
\label{fig:fnl_r_scatter}}
\subfigure[$|\gnl|$ plotted against $|\fnl|$ for all the points in Fig.~\protect\ref{fig:fnl_r_contour}. The green line corresponds to the linear relation $\gnl \sim \fnl$ and the blue line to the quadratic relation $\gnl \sim \fnl^2$.]
{
\includegraphics[width=7 cm]{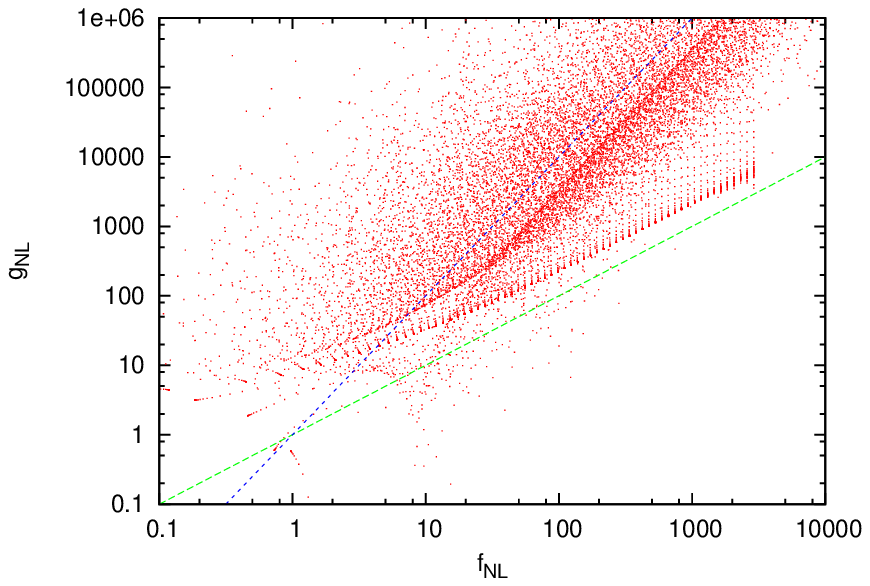}
\label{fig:fnl_r_gnl}}
\caption{The behaviour of $|\fnl|$ and $|\gnl|$ for $n=4$ and $m=10^{-12}$.}
\label{fig:fnl_r}
\end{figure}

In Fig.~\ref{fig:fnl_r_contour} we have plotted the value of
$|\fnl|$ against the initial conditions $r_*$ and $H_*$. The
vertical line in this figure corresponds to a range of parameters, corresponding to
the fixed value of $H_* = 5\times10^{-6}$, for which $|\fnl|$ is
plotted against $\rdec$ in Fig.~\ref{fig:fnl_r_stripe}. Here $\fnl$ can be seen oscillating around the
$1/\rdec$-estimate. Oscillations arise from the derivative terms
present in Eq.~(\ref{f_nl}), which describe the impact of the
self-interaction terms on the dynamics of the curvaton. Most points
give rise to a larger $\fnl$ than one would expect from the estimate $1/\rdec$, however since $\fnl$ actually changes sign, points can
always be found where $\fnl$ is arbitrarily close to zero. In Fig.~\ref{fig:fnl_r_stripe} several different values of $|\fnl|$ correspond to a given point $\rdec$ since different choices of initial conditions can be degenerate yielding the same $\rdec$.

In Fig.~\ref{fig:fnl_r_scatter} we plot again $|\fnl|$ against $\rdec$, but we no longer constrain $H_*$, but instead plot this
for all points in Fig.~\ref{fig:fnl_r_contour}. Again each point
corresponds to a set of parameters producing the observed final
amplitude of the primordial perturbations. As we allow $H_*$ to take
different values, a family of curves is drawn, where each curve is
similar to the curve present in \ref{fig:fnl_r_stripe}, resulting into
the noisy scatter present in the Fig.~\ref{fig:fnl_r_scatter}. It is
also noteworthy that for given fixed value of $\rdec$, there are
multiple sets of parameters which all give the same final amplitude
for the perturbations, but different value of $\fnl$ (and $\gnl$).

In Fig.~\ref{fig:fnl_r_gnl} the value of $|\gnl|$ is plotted against $|\fnl|$. Here three contributions can be clearly
distinguished: The $1/\rdec$-relation arising while in the quadratic regime, the
$1/\rdec^2$-relation in the non-quadratic regime that is due to
self-interactions, and scatter around those lines due to the
oscillations caused by the self-interactions. This scatter can be
understood by considering Fig.~\ref{fig:fnl_r_stripe} where $\fnl$
can be seen oscillating around the analytical estimate. Plotting
$\gnl$ against $\rdec$ would produce a qualitatively similar
plot as Fig.~\ref{fig:fnl_r_stripe} showing the oscillations around
the $1/\rdec$ and $1/\rdec^2$ estimates deriving from the terms in
eq.\ (\ref{f_nl}). However, in general $\fnl$ and $\gnl$ do not
oscillate with the same phase, and hence when plotting $\gnl$ against $\fnl$, scatter is created.

\begin{figure}[!ht]
\centering
\subfigure[$|\fnl|$ plotted against $r_*$ and $H_*$.]
{
\includegraphics[width=7 cm]{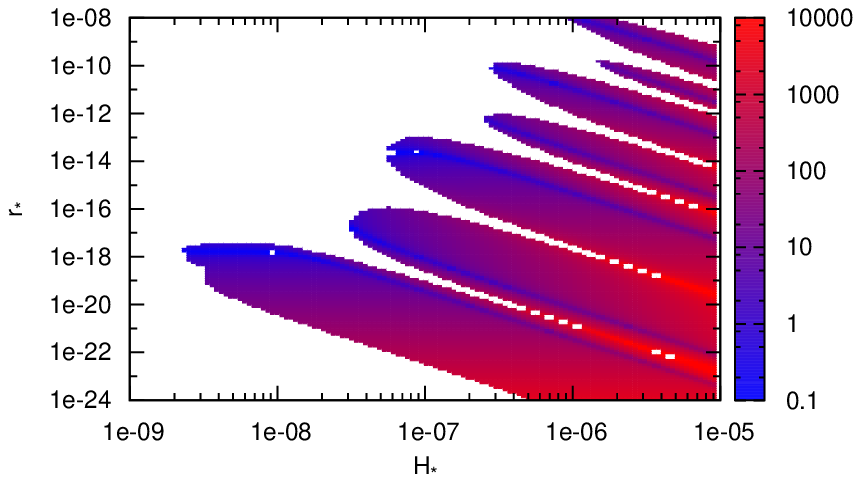}
\label{fig:fnl_gnl_fnl}}
\subfigure[Sign of $\fnl$ plotted against $r_*$ and $H_*$, where the red color corresponds to positive and blue to negative $\fnl$.]
{
\includegraphics[width=7 cm]{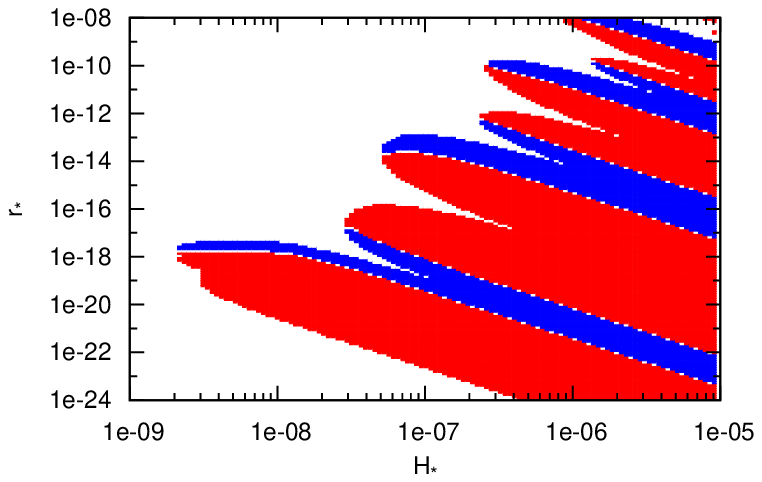}
\label{fig:fnl_gnl_signoffnl}}

\subfigure[$|\gnl|$ plotted against $r_*$ and $H_*$.]
{
\includegraphics[width=7 cm]{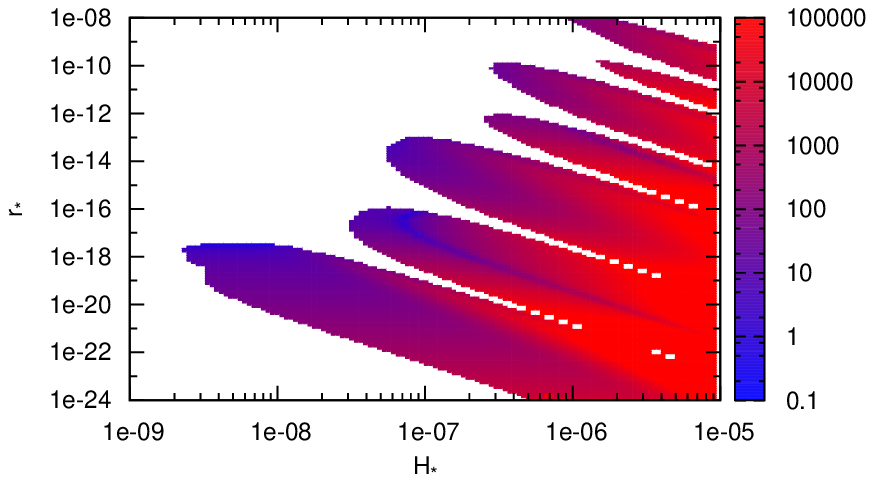}
\label{fig:fnl_gnl_gnl}}
\subfigure[Sign of $\gnl$ plotted against $r_*$ and $H_*$, where red color corresponds to positive and blue to negative $\gnl$.]
{
\includegraphics[height=5 cm]{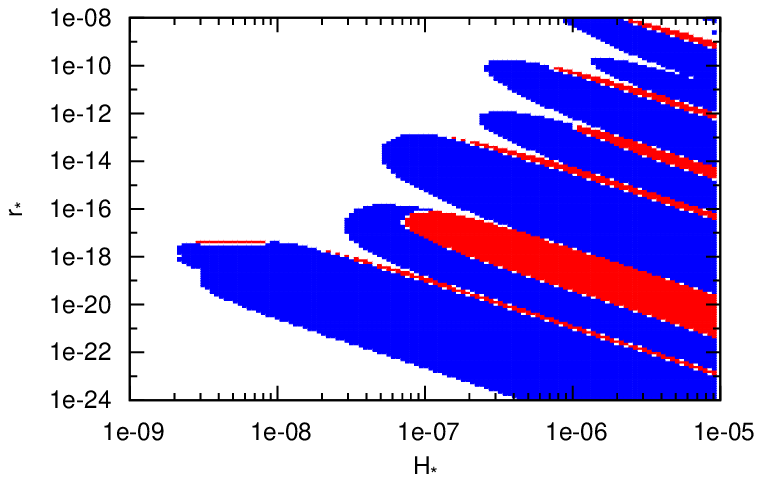}
\label{fig:fnl_gnl_signofgnl}}

\caption{Magnitude of $|\fnl|$ and $|\gnl|$  for $n=4$ and $m=10^{-12}$.}
\label{fig:fnl_gnl}
\end{figure}

In Fig.~\ref{fig:fnl_gnl_fnl} - \ref{fig:fnl_gnl_signofgnl} the
magnitudes and signs of $\fnl$ and $\gnl$ are plotted when scanned
through different initial conditions. Here again the oscillatory
features can be clearly distinguihed as both $\fnl$ and $\gnl$
oscillate in the regime initially dominated by the non-quadratic
interaction. Moreover it is worth emphasizing that not only does the
absolute magnitude of $\fnl$ and $\gnl$ show oscillatory behaviour in
this regime, but also their signs change along the oscillations as
shown in Fig.~\ref{fig:fnl_gnl_signoffnl} and
\ref{fig:fnl_gnl_signofgnl}. Futhermore the oscillations of $\fnl$ and
$\gnl$ have different periods and phases, i.e.\ the zeros of $\fnl$
and $\gnl$ are not related in a simple fashion.

In Eq.\ (\ref{eq:fnlprime}) we presented a relation between the derivative of $\fnl$ with respect to $\sigma_*$ and the values of $\gnl$ and $\fnl$. If $\fnl=0$, this takes a particularly simple form, $\fnl'=9/5N'\gnl$, which can be clearly seen in Fig.~\ref{fig:fnl_gnl_gnl}.

Note that there are two different sources of non-Gaussianity: one is the subdominance of the curvaton at the time of decay $\rdec$, while the other is just the non-linear evolution of curvaton perturbations, encoded in the function $\sigma_{\rm osc}$ in Eqs.~(\ref{f_nl}) and (\ref{g_nl}). Even if $\rdec \sim 1$, large non-Gaussianity can be generated by the evolution of the curvaton. For example, from Eq.~(\ref{f_nl}) we see that if $\sigma_{\rm osc}'\rightarrow 0$, $\fnl$ can be very large even though $\rdec \sim 1$. This can be understood qualitatively by looking at the expression for the linear part of the curvature perturbation, $\zeta\sim \rdec(\sigma_{\rm osc}'/\sigma_{\rm osc})\delta\sigma_{*}$ \cite{LR}. If $\sigma_{\rm osc}' \rightarrow 0$, we see that we need to increase $\delta \sigma_{*}$ to keep $\zeta\sim 10^{-5}$. Therefore, the higher order terms become significant in this limit generating large non-Gaussianities, just like in the limit $\rdec \rightarrow 0$.

\subsection{Allowed regions of parameter space}

\begin{figure}[!ht]
\centering
\subfigure[$n=0$ and $m=10^{-8}$] {\includegraphics[width=7 cm]{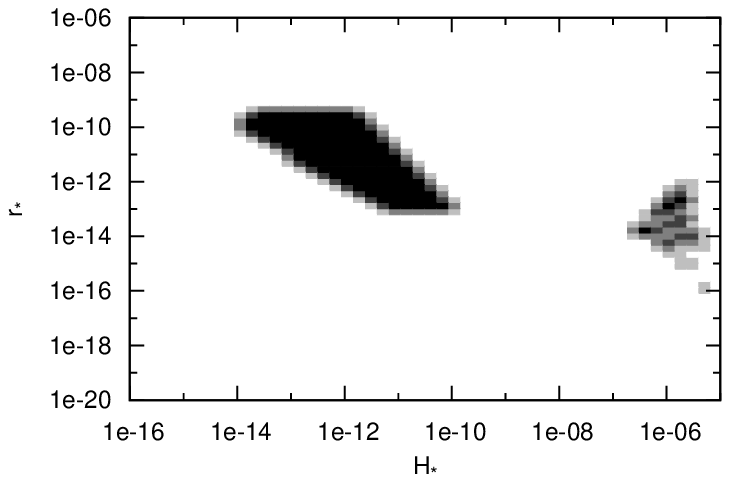}}
\subfigure[$n=0$ and $m=10^{-10}$] {\includegraphics[width=7 cm]{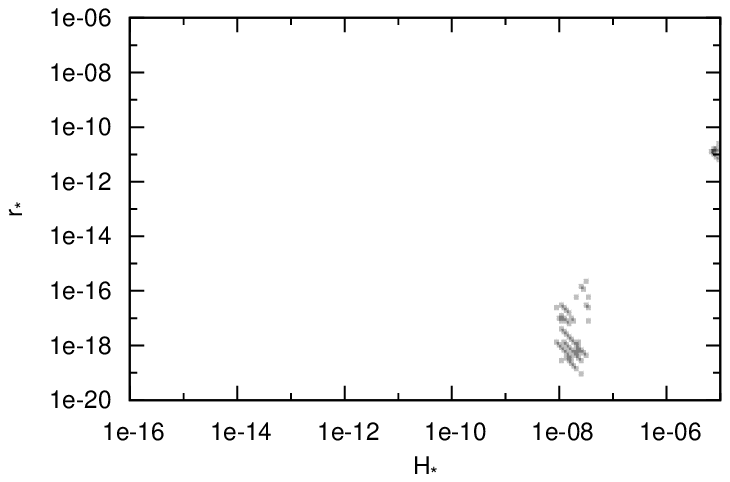}}

\subfigure[$n=6$ and $m=10^{-10}$] {\includegraphics[width=7 cm]{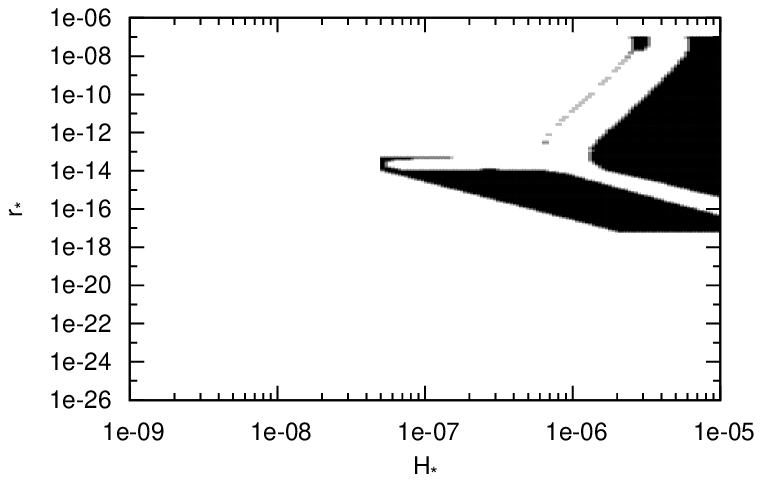}}
\subfigure[$n=6$ and $m=10^{-12}$] {\includegraphics[width=7 cm]{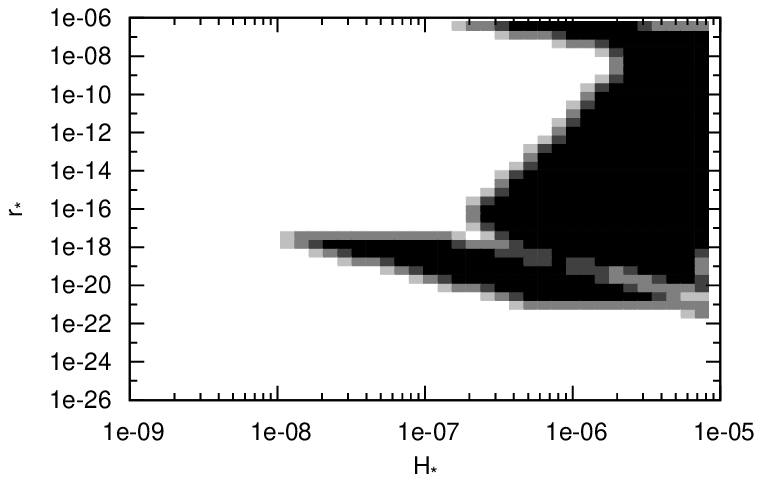}}
\caption{Dark areas correspond to the allowed areas in the parameter space with $-9 < \fnl < 111$ and
$-3.5\times10^5 < \gnl < 8.2\times10^5$ for $n=0$ and $n=6$.}
\label{fig:resultsn0n6}
\end{figure}

\begin{figure}[!ht]
\centering
\subfigure[$n=2$ and $m=10^{-8}$] {\includegraphics[width=7 cm]{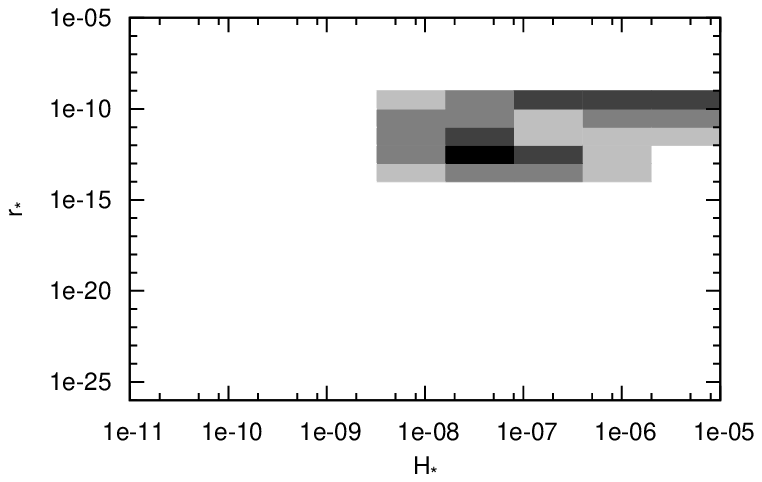}}
\subfigure[$n=2$ and $m=10^{-10}$] {\includegraphics[width=7 cm]{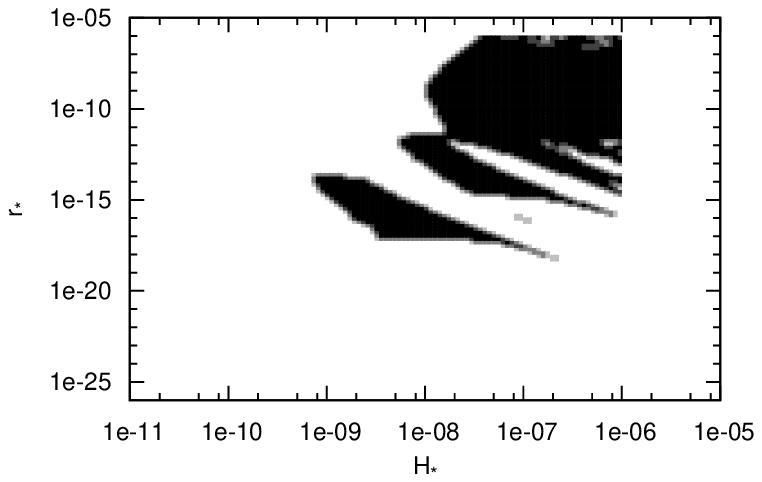}}

\subfigure[$n=2$ and $m=10^{-12}$] {\includegraphics[width=7 cm]{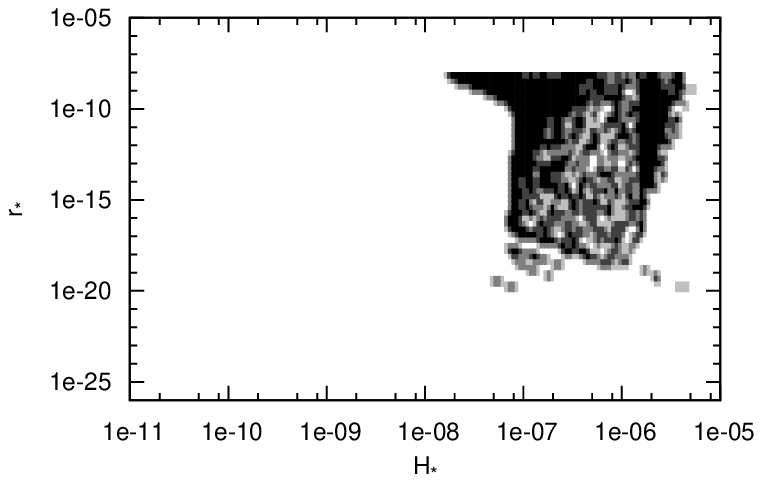}}
\subfigure[$n=2$ and $m=10^{-14}$] {\includegraphics[width=7 cm]{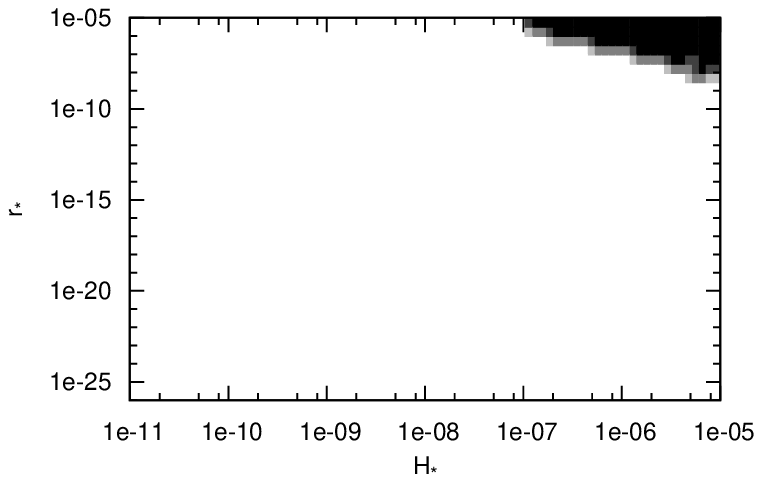}}
\caption{Dark areas correspond to the allowed areas in the parameter space where $-9 < \fnl < 111$ and $-3.5\times10^5 < \gnl < 8.2\times10^5$ for $n=2$.}
\label{fig:resultsn2}
\end{figure}

\begin{figure}[!ht]
\centering
\subfigure[$n=4$ and $m=10^{-8}$] {\includegraphics[width=7 cm]{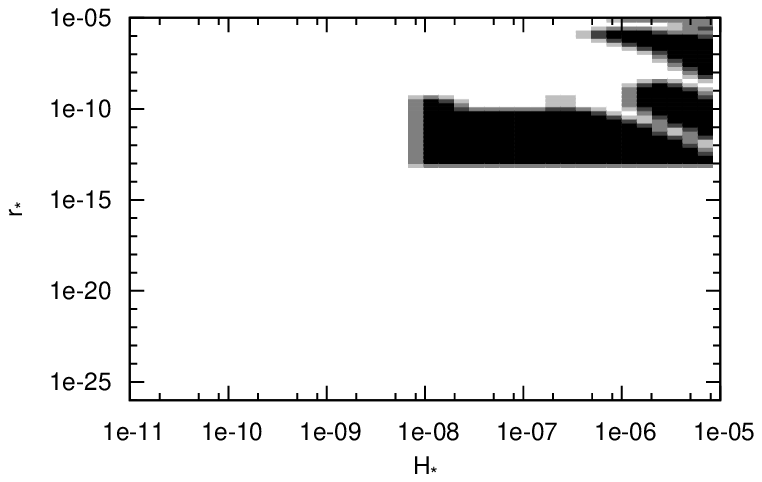}}
\subfigure[$n=4$ and $m=10^{-10}$] {\includegraphics[width=7 cm]{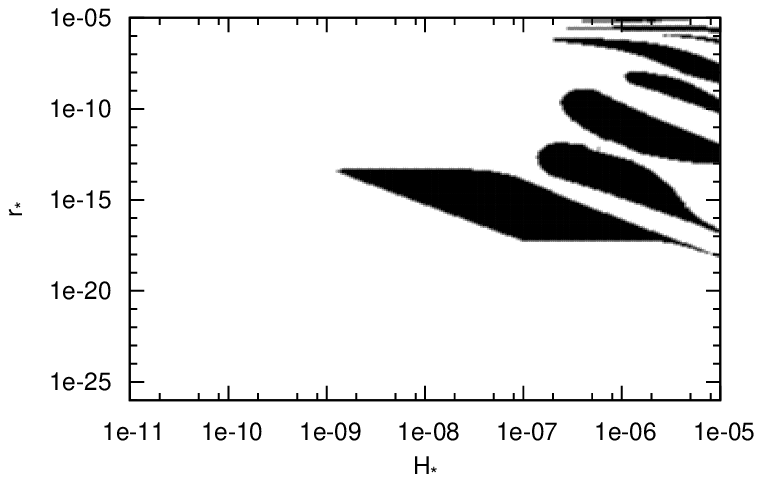}}

\subfigure[$n=4$ and $m=10^{-12}$] {\includegraphics[width=7 cm]{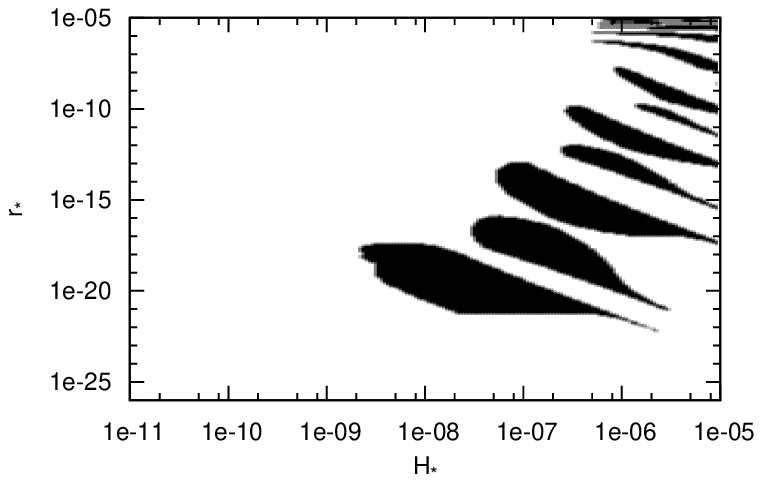}}
\subfigure[$n=4$ and $m=10^{-14}$] {\includegraphics[width=7 cm]{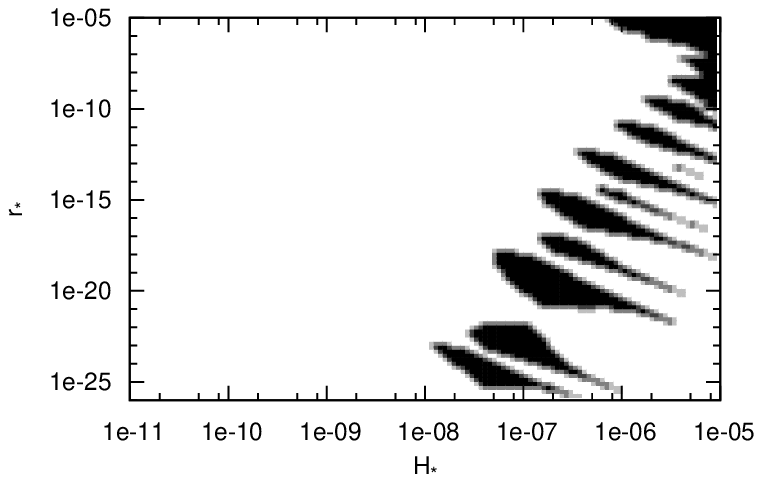}}
\caption{Dark areas correspond to the allowed areas in the parameter space where $-9 < \fnl < 111$ and $-3.5\times10^5 < \gnl < 8.2\times10^5$ for $n=4$.}
\label{fig:resultsn4}
\end{figure}

In Figs.~\ref{fig:resultsn0n6} - \ref{fig:resultsn4} we have plotted
the points in the parameter space which are compatible with
observations of the primordial perturbations. These points give rise to
the observed amplitude of the perturbations while producing $\fnl$ and
$\gnl$ which are within the current observable limits.

The range of the initial conditions $H_*$ and $r_*$ has been chosen so
that all interesting features should be within the plots. $H_*$ is
also bounded from above, $H_* \lesssim 10^{-5}$, to prevent the excessive
production of primordial tensor modes and at least an order of magnitude smaller to keep the inflaton perturbations negligible.

Limits for $\fnl$ are given by the WMAP 5-year data \cite{wmap}, $-9 <
\fnl < 111$. Although in \cite{Smith:2009jr} a more stringent constraint
for $\fnl$ is given as $ - 4 < \fnl <80$, we conservatively use
the limit provided in \cite{wmap}. Regarding the limits for $\gnl$,
we require that $\gnl$ is the range $-3.5\times10^5 < \gnl <
8.2\times10^5$ as given in \cite{uros}. Note that these limits have been derived assuming
$\fnl \sim 0$, which here is not the case in general. However, the bounds for
$\fnl$ seem to be much more constraining in our case, and relaxing the
limits for $\gnl$ even by an order of magnitude would not enlarge the
allowed area of the parameter space significantly.
Thus we adopt this limit for reference purposes\footnote{
Recently, the authors of \cite{Vielva:2009jz} have obtained
limits on $\gnl$ without assuming $\fnl = 0$ by using N-point probability
distribution function. However, their constraint on $\gnl$
is similar to that of \cite{uros}.}.

\begin{figure}[!ht]
\centering
\includegraphics[height=9cm]{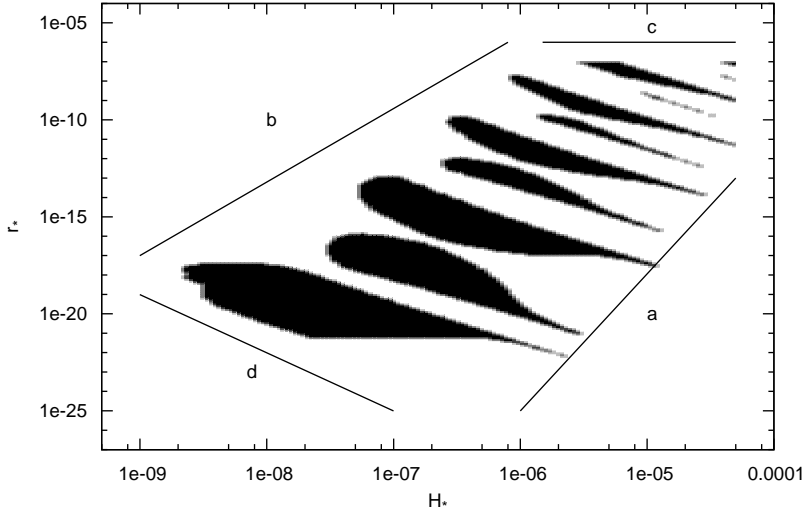}
\caption{A schematical illustration of the different cuts limiting the allowed area in the parameter space.}
\label{fig:lines}
\end{figure}

In Fig.~\ref{fig:lines} we give an illustration of the different features limiting the allowed
area of the parameter space. The observational
limits for $\fnl$ and $\gnl$ constrain the allowed area in the very
subdominant regions of the parameter space, as depicted in Fig.
\ref{fig:lines} by the line \emph{a}. Other constraints shown in
Fig.~\ref{fig:lines} arise from the internal consistency of the specific curvaton scenario
studied in the present paper. The bound \emph{b} is obtained because otherwise the
initial perturbations would be too small to produce the observed
amplitude. The bound \emph{c} reflects
the requirement that the curvaton should be massless, or $V'' < H_*^2$,
which is necessary for the generation of curvaton perturbations during
inflation. Because of the subdominance of the curvaton, the realistic
bound should arguably be a few orders of magnitude tighter. However,
a change of an order of magnitude moves the actual cut by a very
small amount in the log-log plots.

Finally, the bound \emph{d}
guarantees the absence of the isocurvature modes in dark matter perturbations and
corresponds to the limit on the curvaton decay width given in
(\ref{eq:isocurvature}).

It should be noted, that limits \emph{b}, \emph{c} and \emph{d} in
Fig.~\ref{fig:lines} were already present in
\cite{Enqvist:2009zf}, so that the non-Gaussianty limit \emph{a} is
the only new limiting ingredient provided by the current work on the space of parameters.

As explained in the previous sections, in the regions where the
quadratic part of the potential dominates already initially, the
dynamics are essentially linear, and the simple results
apply. As a consequence, $\sigma_{\mathrm{osc}}$ dependence in Eqs. (\ref{f_nl})
and (\ref{g_nl}) disappears. Therefore these regions are characterized
by the smooth continuous allowed area as shown in Figs.
\ref{fig:resultsn0n6}~-~\ref{fig:resultsn4}, which can be found in the
lower left are in the plots. The total area of the allowed region depends on
the values of $m$ and $n$; e.g.\ for $n=4$ and $m=10^{-8}$ this
quadratic area is much larger than for, say,  $n=4$ and $m=10^{-14}$.

As mentioned previously, $n=0$ and $n=6$ do not display
extended oscillatory regions. Hence the plots in Fig.
\ref{fig:resultsn0n6} remain smooth also in the regime where the self-interaction
dominates.

The cases $n=2$ and $n=4$ are however characterized by significant
oscillatory features, as discussed in the previous sections. Due to
these oscillations, the allowed region consists of isolated patches in
the interaction dominated regime. As can be seen in the figures, the
size of these patches decreases as we decrease the (bare) mass $m$,
since this tends to make the curvaton more subdominant at the time of decay, decreasing the
viable area.

%\footnote{This is due to the fact that by decreasing the mass decreases the time that the curvaton spends in the quadratic regime where the energy density of the curvaton increases compared to the background.}

\section{Discussion}
\label{yakyak}

It seems plausible that the curvaton, like any other scalar field, has
some self-interactions. These self-interactions can have a profound effect on the dynamical
evolution of the curvaton field and its perturbations, as was
discussed in \cite{Enqvist:2009zf} (and in \cite{kesn,kett})), where we studied the amplitude of
the curvature perturbation in self-interacting curvaton models defined by Eq.~(\ref{V}).  In the present paper
we have focussed on the non-Gaussianities of the curvature
perturbations by computing the non-linearity parameters $\fnl$ and $\gnl$ for all the model parameters for which $\zeta\sim 10^{-5}$.

When the curvaton has some self-interactions,
the non-Gaussian statistics of the curvature perturbation can be quite different from that produced in the simplest model with a quadratic potential.
In the quadratic curvaton model, the magnitude of $\fnl$ in the limit $\rdec \ll 1$ is determined by the curvaton energy density at the time of its decay, $\fnl \sim 1/\rdec$. However, the prediction of $\fnl$ can significantly deviate from this simple estimate if the curvaton has non-renormalizable self-interactions. As seen from Fig.~\ref{fig:fnl_r_scatter}, for such models the values of $\fnl$ scatter around the estimate $1/\rdec$ and typically end up being slightly larger than in the quadratic case. Thus a very subdominant curvaton
is not favoured because it tends to yield a value of $\fnl$ which is in excess of the
observational bounds
$| \fnl | \lesssim 100$  \cite{wmap,Smith:2009jr}.

However, it is interesting to note that even if $\rdec \ll 1$, there exists regions in the parameter space with $|\fnl| < \mathcal{O}(1)$.
This is because the value of $\fnl$ oscillates and changes its sign,
as is illustrated in Fig.~\ref{fig:fnl_r_contour} for $n=4$.  However, even if $\fnl < \mathcal{O}(1)$,
the non-linearity parameter for the trispectrum $\gnl$ can be
very large, as can be seen in Fig.~\ref{fig:fnl_r_gnl}.
In these regions the self-interacting curvaton scenario gives rise to a rather non-trivial non-Gaussian statistics characterized by a large trispectrum and a vanishing bispectrum. Such a situation, discussed already in \cite{kett}, appears to be rather generic in self-interacting curvaton models, and possible for a wide, albeit restricted, range of model parameters.

Another interesting feature of the curvaton model with self-interactions is that large non-Gaussianities $|\fnl| \gg \mathcal{O}(1)$ can be generated even if the curvaton dominates the energy density at the time of its decay, $\rdec \sim 1$. Recently in \cite{Nakayama:2009ce} it was shown that for $\rdec\sim1$, the entropy production at the curvaton decay can leave an imprint on the spectrum of primordial graviational waves, which in principle could be observable. If the curvaton had no self-interactions, such a signal would imply that no large non-Gaussianity could be generated by the curvaton fluctuations. This is clearly not the case when the self-interactions are included, which serves to demonstrate thet the self-interactions can significantly affect the generic features of the curvaton scenario.

Another interesting feature that is clearly visible in Fig.~\ref{fig:fnl_r_gnl} is the
breakdown of the relation of $\gnl = -(10/3) \fnl$ which holds true for
the quadratic potential. For
the self-interacting curvaton
a large number of points, each corresponding to an allowed set of parameter values,
can be seen to fall into the region with $| \gnl | > |\fnl |$.
If the interactions are small compared to the quadratic part, the linear relation $\gnl = -(10/3) \fnl$ gets replaced by $\gnl \sim -\fnl^2$\,\cite{kett}.
However, when the self-interaction term dominates, both of the above relations can be violated,
as is seen from Fig.~\ref{fig:fnl_r_gnl}.
There is nevertheless a tendency for the points to be concentrated
around the lines  $| \gnl | \sim |\fnl|$ and
$| \gnl | \sim |\fnl^2 |$.

We should also like to point out that in the quadratic case with $|\fnl| \gg 1$, the signs of $\fnl$ and $\gnl$
are respectively positive and negative. In contrast,
in the cases studied here,
 the signs can well be the same.
This feature provides  an obvious departure from the quadratic case
and could offer an experimental possibility to constrain the curvaton self-interaction.

It is evident from the figures presented in Sect.~\ref{nonrenormpot} that  the self-interacting curvaton model
provides us a rich tapestry of features, constrained in a highly non-trivial way by
the observational limits on non-Gaussianity.
As we have argued here, in the presence of self-interactions the relative signs of $\fnl$ and $\gnl$ and
the functional relation between them is typically modified from the quadratic case. Thus the
non-linearity parameters taken together, in possible conjunction of other cosmological observables such as tensor perturbations,
may offer the best prospects for constraining the physical properties of the curvaton.

\acknowledgments

This work is supported in part by the Grant-in-Aid for Scientific
Research from the Ministry of Education, Science, Sports, and
Culture of Japan No.\,19740145 (T.T.), by the EU 6th Framework Marie
Curie Research and Training network "UniverseNet"
(MRTN-CT-2006-035863), by the Academy of Finland grants 114419
(K.E.) and 130265 (S.N.) and by the Magnus Ehrnrooth Foundation
(O.T.).

\end{document}